\documentclass[noshowpacs,preprint,aps,prb,citeautoscript]{revtex4}

\usepackage{amsmath,bm}
\usepackage{graphicx}
\usepackage{color}
\usepackage{epsfig}
\usepackage{times,psfrag}
\usepackage{array}
\usepackage{tikz}
\usepackage{xspace}

\begin{document}


\title{Morphological analysis of chiral rod clusters \\ 
from a coarse-grained single-site chiral potential } 



\author{B. J. Sutherland}
\affiliation{Physical \& Theoretical Chemistry Laboratory, South Parks Road, Oxford OX1 3QZ, United Kingdom}
\author{S.\ W.\ Olesen}
\affiliation{Harvard T.\ H.\ Chan School of Public Health, 677 Huntington Avenue, Boston, MA 02115, United States of America}
\author{H.\ Kusumaatmaja}
\email[Email: ]{halim.kusumaatmaja@durham.ac.uk}
\affiliation{Department of Physics, University of Durham, South Road, Durham, DH1 3LE, United Kingdom}
\author{J.\ W.\ R.\ Morgan}
\affiliation{University Chemical Laboratories, University of Cambridge, Lensfield Road, Cambridge, CB2 1EW, United Kingdom}
\author{D.\ J.\ Wales}
\email[Email: ]{dw34@cam.ac.uk}
\affiliation{University Chemical Laboratories, University of Cambridge, Lensfield Road, Cambridge, CB2 1EW, United Kingdom}

\begin{abstract}
We present a coarse-grained single-site potential for simulating chiral interactions, with adjustable strength,
handedness, and preferred twist angle. As an application, we perform basin-hopping global optimisation to predict the
favoured geometries for clusters of chiral rods. The morphology phase diagram based upon these predictions has four
distinct families, including previously reported structures for potentials that introduce chirality based on shape,
such as membranes and helices. The transition between
these two configurations reproduces some key features of experimental results for {\it{fd}} bacteriophage. The
potential is computationally inexpensive, intuitive, and versatile;  we expect it will be useful for
large scale simulations of chiral molecules. For chiral particles confined in a cylindrical container we
reproduce the behaviour observed for fusilli pasta in a jar. Hence this chiropole potential has the capability
to provide insight into structures on both macroscopic and molecular length scales.
\end{abstract}

\maketitle

\section{Introduction}

Chirality is ubiquitous and important in nature. Many key biological molecules,
such as proteins and nucleic acids, are chiral. Their homochirality \cite{Blackmond10},
and the role chirality plays in interactions at the molecular, cellular, and multicellular
levels, remain important questions crucial for morphogenesis \cite{WanRPTZGVN11,McsheeneB11,Wada12},
and thus our understanding of life. In the context of materials science, chiral structures,
ranging from chiral liquid crystal phases \cite{deGennes71,ChiralBook,YuanMSTS18} to colloidal clusters
\cite{ZerroukiBPCB08,ChakrabartiW11} and nanotubes \cite{DingHY09,LiuDY10,TeichSIT12},
have also attracted great interest, due to their unusual optical, mechanical, and chemical properties.
The cholesteric liquid crystal phase has attracted recent attention due to the
challenges of efficient modelling.\cite{TortoraD17a,TortoraD17b,TortoraD18}

Both in biology and materials science, one of the most pertinent questions is to understand
how chirality is transmitted across length scales.\cite{DussiBRD15} A number of studies have shown that chiral
structures do not actually require chiral subunits; they can arise from achiral building blocks,
\cite{ChakrabartiFW09,YanHE08,EdlundLJ12} or even chiral building
blocks of opposite handedness.\cite{OlesenFCW13,FrezzaFKGC14} Conversely, given a chiral building block,
it is useful to understand and predict the large-scale functional superstructures that
may self-assemble, and whether
we can control the corresponding morphologies and physical properties,
particularly if these structures are themselves chiral.

We are primarily interested in the second issue in this contribution. Our motivation is to
describe a computationally inexpensive and intuitive coarse-grained chiral potential,
thus facilitating future large-scale simulations of chiral assemblies. The chiropole interaction
potential we describe is single-site. It is easier to evaluate than for models where chirality is
introduced through the shape of the molecule, which requires multisite interaction potentials.
\cite{YanHE08,YanHE09,HorschZG06,KolliCFG16,CinacchiFGK17}
\textcolor{black}{Unlike a similar single-site potential proposed by van der Meer \textit{et al.}\cite{MeerVDY76}\ for lattice simulations of liquid crystals, our potential is polar in that it distinguishes the two ends of the molecule.}
We can adjust the strength, handedness,
and preferred twist angle independently, with the degree of chirality controlled by the twist angle between two chiral molecules.
Hence, we hope the single-site chiral potential could prove to be an intuitive and versatile
representation.\cite{MemmerKS93}

For two rod-shaped rigid bodies, chiral interactions can be modelled by enforcing
a relative twist of the two long axes.\cite{HarrisKL99} A familiar example is
provided by fusilli pasta contained within a jar, where the spirals near the centre align almost
vertically, but towards the edges a significant twist is observed.\cite{SchallerB12} The propensity for
local twisting interferes with the ability to achieve close packing.

As an application, we {\color{black} exploit} our potential by running global optimisation for
structure predictions of clusters of chiral rods, over a wide range of simulation
parameters. We observe many morphologies that have been previously reported in experiments
and/or computer simulations using more complex potentials.\cite{YanHE08} In particular, we compare to
recent experiments on {\it{fd}} bacteriophage \cite{GibaudBZHWYBO12}, which reported morphologies
including membranes, twisted ribbons, and rings by tuning the chirality. These results
suggest that the chiropole potential successfully captures key aspects of the interparticle
forces, enabling us to extract the minimal conditions for particular morphologies
to emerge. 

Our report is organised as follows. We introduce the single-site chiral potential in the
next section. We then summarise the basin-hopping algorithm \cite{LiS87,WalesS99,WalesD97}
for reference, and define a spherocylinder potential \cite{AllenEFM93,VegaL94} in section III.
The spherocylinder potential is used to give the chiral particles a rod shape, but in principle any other
potential, from Lennard-Jones \cite{LennardJones24} or Morse \cite{Morse29} for a
spherical shape, to Gay-Berne \cite{GayB81} or Paramonov-Yaliraki \cite{ParamonovY05}
for an ellipsoidal form, could be used. We discuss the global optimisation results for clusters of
chiral rods in section IV, and finally we summarise the most important results and discuss
avenues for future research in section V. 

\section{The single-site chiral potential}

Our single-site chiral potential is given by
\begin{equation} \label{eq:chiro_pot}
    U^c_{ij} = -\frac{\mu^2 \sigma_0^3}{r_{ij}^3} \left[ \cos\alpha \left(\hat{\bm{\mu}}_i \cdot \hat{\bm{\mu}}_j\right) +
\sin\alpha \left(\hat{\bm{\mu}}_i \times \hat{\bm{\mu}}_j\right) \cdot \hat{\mathbf{r}}_{ij}\right],
\end{equation}
where $\mu$ is the interaction strength, $\sigma_0$ defines the length scale, and $r_{ij}$ is the
intercentre distance between two chiral particles $i$ and $j$. $\hat{\bm{\mu}}_i$ and $\hat{\bm{\mu}}_j$ describe
the orientations of the chiral poles, and the angle $\alpha$ determines preferred twist angle.

\begin{figure}
\centering
\includegraphics[width=0.9\textwidth]{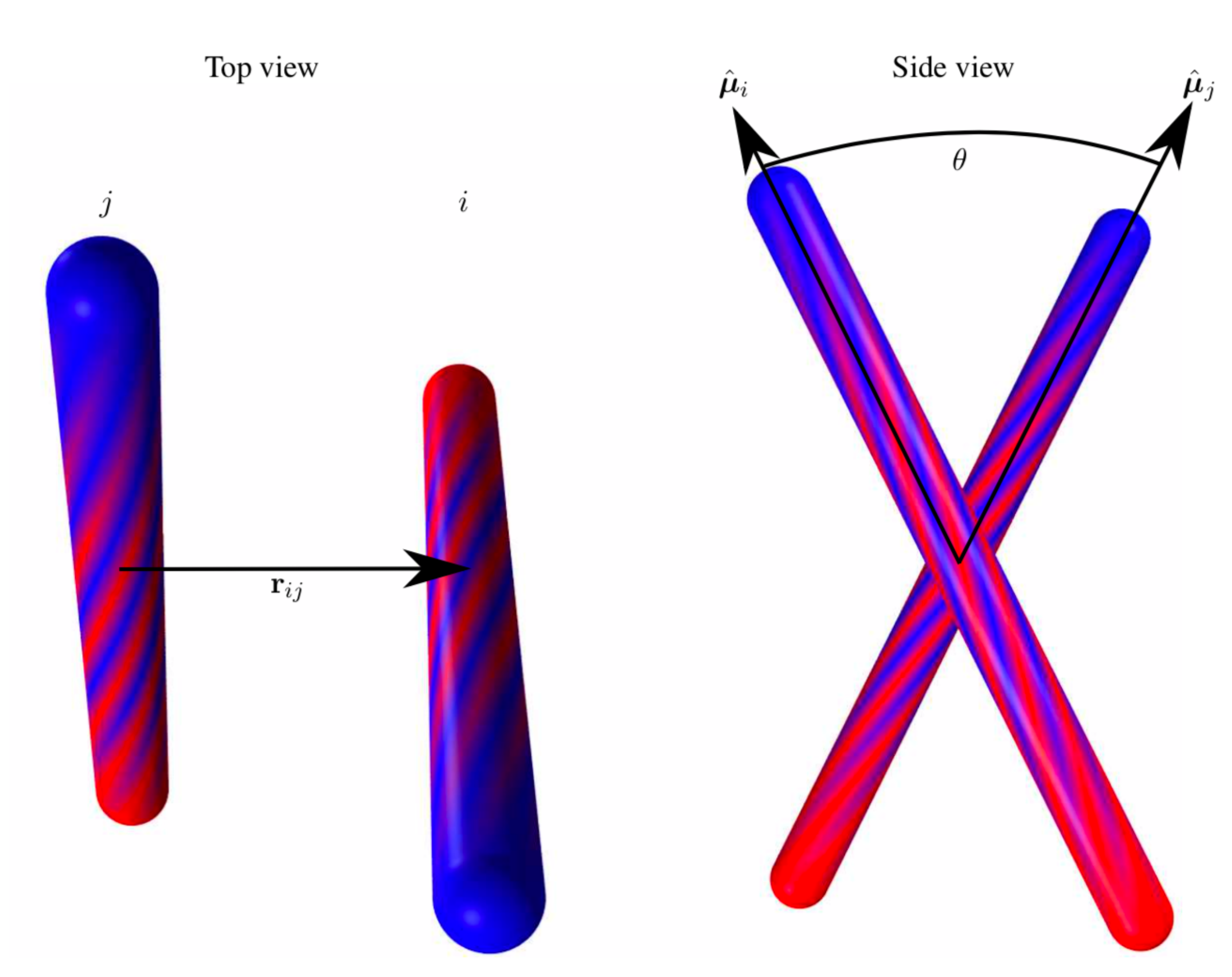}
\caption{The configuration for a dimer of chiral poles, showing the intercentre vector $\mathbf{r}_{ij}$,
the orientations of the chiral poles $\hat{\bm{\mu}}_{i}$ and $\hat{\bm{\mu}}_{j}$, and the angle separating
the chiral poles $\theta$. At the minimum energy for the chiral interaction, $\theta = \alpha$.}
\label{schematic}
\end{figure}

The $\left(\hat{\bm{\mu}}_i \cdot \hat{\bm{\mu}}_j\right)$ term is similar to the first term in a dipolar interaction.
It favours orientations in which $\hat{\bm{\mu}}_i$ and $\hat{\bm{\mu}}_j$ are parallel, since $U^c$ has an overall
minus sign. The $\left(\hat{\bm{\mu}}_i \times \hat{\bm{\mu}}_j\right) \cdot \hat{\mathbf{r}}_{ij}$ term, on the other
hand, favours orientations in which $\hat{\bm{\mu}}_i$ and $\hat{\bm{\mu}}_i$ are at right angles and
$\left(\hat{\bm{\mu}}_i \times \hat{\bm{\mu}}_j\right)$ is parallel to $\hat{\mathbf{r}}_{ij}$,
where $\hat{\mathbf{r}}_{ij} = \frac{\mathbf{r}_{i} - \mathbf{r}_{j}}{r_{ij}}$ and $\mathbf{r}_{i}$ is the
centre of mass of particle $i$. Let us now
concentrate on a dimer configuration. If the distance $r_{ij}$ is fixed, see figure \ref{schematic},
then $\left(\hat{\bm{\mu}}_i \cdot \hat{\bm{\mu}}_j\right) = \cos{\theta}$ and 
$\left(\hat{\bm{\mu}}_i \times \hat{\bm{\mu}}_j\right) \cdot \hat{\mathbf{r}}_{ij} = \sin{\theta}$,
where $\theta$ is the angle between the chiral poles. The potential is then proportional to
\begin{equation}
\cos{\alpha} \cos{\theta} + \sin{\alpha} \sin{\theta} = \cos{(\alpha-\theta)} \, ,
\end{equation}
and it is minimised when $\theta = \alpha$. For $\alpha = 0$, the interaction is achiral,
and the poles prefer to be aligned. For $\alpha = \pi$, the interaction is still achiral, but the poles
prefer to be antiparallel. The interaction is {\color{black} chiral for $\alpha = \pi/2$ as the two ends of the pole are
distinguishable.} We further note that reversing the sign of $\alpha$ reverses the handedness of the
chiral interaction.

{\color{black} The potential is attractive when the particles are close to alignment. A purely repulsive potential,
with minimum repulsion when the particles are aligned, could also be considered. Since the chiropole potential will
be added to some other function defining the shape of the particles, and must be attractive overall
for particles to assemble, these different choices merely amount to shifting the magnitude of the overall potential
and will not qualitatively affect the behaviour. Choosing the simplest possible representation that could model the
chiral nature of interactions was our primary concern.}

{\color{black} The chiropole potential is similar to that proposed by van der Meer \textit{et al.\ }\cite{MeerVDY76} in the context of liquid crystals,
which has since been used in lattice simulations.\cite{MemmerF03,BiagioSERZ18} The corresponding function is}
\begin{equation} \label{eq:Meer_pot}
    U^c_{ij} = -\frac{\mu^2 \sigma_0^3}{r_{ij}^3} \left[ \cos\alpha \left(\hat{\bm{\mu}}_i \cdot \hat{\bm{\mu}}_j\right)^{2} +
\sin\alpha \left(\hat{\bm{\mu}}_i \cdot \hat{\bm{\mu}}_j\right) \left(\hat{\bm{\mu}}_i \times \hat{\bm{\mu}}_j\right) \cdot \hat{\mathbf{r}}_{ij}\right],
\end{equation}
{\color{black}The inclusion of the extra $\left(\hat{\bm{\mu}}_i \cdot \hat{\bm{\mu}}_j\right)$ term makes the potential symmetric
with respect to inverting the direction of a particle. The maximum supportable preferred angle is therefore $\frac{\pi}{2}$ and
the point group symmetry of a particle is $D_{\infty}$. Our
potential is not symmetric on inverting the direction of a particle, producing point group $C_{\infty}$, and it is suitable for modelling
particles in which the two ends are not identical. \textit{fd} bacteriophage is one such system.}\cite{GailusR94}

\section{Methods}

\subsection{Spherocylinders}

As for the dipolar interactions, the potential defined in Eq.\ \eqref{eq:chiro_pot} has a singularity at $r_{ij} = 0$.
Building blocks used in simulations should therefore have repulsive cores added to prevent sites from falling into
this singularity. For example, a Lennard-Jones \cite{LennardJones24} or Morse \cite{Morse29} potential
could be used for spherical particles, while Gay-Berne \cite{GayB81} and Paramonov-Yaliraki \cite{ParamonovY05}
potentials are suitable options for ellipsoids. Here we are particularly motivated by recent experimental results
for the {\it{fd}} bacteriophage \cite{GibaudBZHWYBO12}. The {\it{fd}} virus is 880 nm long and has a diameter of 6.6 nm,
yielding an aspect ratio of over 100. We therefore use spherocylinders \cite{AllenEFM93,VegaL94} to give the
chiral particles a rod shape. The corresponding additional potential is then
\begin{equation}
U^{r}_{ij} = 4 \epsilon_{r} \left[ \left(\frac{\sigma_r}{d}\right)^{12} - \left(\frac{\sigma_r}{d}\right)^{6} \right] \, ,
\label{eq:spherocylinder}
\end{equation}
where $\epsilon_{r}$ and $\sigma_r$ provide the energy and length scales for the rod/core interactions,
and $d$ is the distance of closest approach between the two rods, computed using the algorithm of Vega and Lago
\cite{VegaL94}, described in the appendix.
A further parameter, $L$, sets the aspect ratio of the spherocylinder and is equal to the distance from the centre of the
rod to an end point. $L$ appears in the computation of $d$: see the appendix.

\subsection{The Pasta Jar}

To model the pasta jar phenomenon described above,\cite{SchallerB12} a cylindrical container was introduced,
with height $h$ and radius $R$, centred at the origin and with the long axis aligned
along the $z$ direction. To repel rods from the container walls, the smallest
distance $r_{\rm min}$ was calculated between both ends of each rod and each of the three surfaces
of the cylinder. A repulsive $r_{\rm min}^{12}$ potential was then calculated for each of these six
distances to provide a sharp cut-off close to the container walls. For particle $i$, the
contributions are

\begin{eqnarray}
    U_{i, \mathrm{top}}^{\pm} &=& \epsilon_{w} \left(\frac{\sigma_{w}}{\frac{h}{2} - \left(\mathbf{r}_{i} \pm L \hat{\bm{\mu}}_{i}\right)_{z}}\right)^{12} \\
    U_{i, \mathrm{bottom}}^{\pm} &=& \epsilon_{w} \left(\frac{\sigma_{w}}{\frac{h}{2} + \left(\mathbf{r}_{i} \pm L \hat{\bm{\mu}}_{i}\right)_{z}}\right)^{12} \\
    U_{i, \mathrm{curve}}^{\pm} &=& \epsilon_{w} \left(\frac{\sigma_{w}}{R - \sqrt{ \left(\mathbf{r}_{i} \pm L \hat{\bm{\mu}}_{i}\right)_{x}^{2}
        + \left(\mathbf{r}_{i} \pm L \hat{\bm{\mu}}_{i}\right)_{y}^{2}}}\right)^{12},
\end{eqnarray}
where $\sigma_{w}$ and $\epsilon_{w}$ are scaling parameters for the length and energy of the container
repulsion, respectively, which we set equal to $\sigma_{r}$ and $\epsilon_{r}$. The $x$, $y$ and $z$ subscripts represent Cartesian components of the
relevant vectors, and the $\pm$ symbols apply to the top and bottom of each rod.

To prevent particles from escaping the container, after any step that would bring
either of the end positions outside of the boundary, the rods in question are progressively
scaled towards the centre and aligned close to the $z$ axis, and this process
is iterated until the rod is entirely within the cylinder. If a rod escapes the cylinder
during energy minimisation then that step is rejected. Attention was focused
on parameters with {\color{black}$\alpha < \frac{\displaystyle \pi}{\displaystyle 2}$}, where the rods should prefer to arrange parallel to
each other rather than antiparallel, so that the runs could be started with the rods
aligned close to the positive $z$ direction of the cylinder, from which it was expected
that convergence to the global minimum would be faster.
The results presented below were calculated with a value of unity for all energy and
length scales, and $L = 6$ was chosen to provide a moderate aspect ratio.

{\color{black} As fusilli pasta has approximate point group symmetry $D_{\infty}$, with
the ends being equivalent, we compared the results for the chiropole potential with
the van der Meer potential.}

\subsection{Basin-hopping}

We have located the putative global minima of clusters of chiral spherocylinders using basin-hopping
global optimisation \cite{LiS87,WalesS99,WalesD97}. In this approach, after
a trial move based on coordinate perturbations is proposed, it is followed by an energy minimisation; the move is then accepted or rejected
based upon the change in energy for the local minima. A simple, yet effective, approach is to use a
Metropolis acceptance criterion: a step is accepted if $U_{\mathrm{new}} < U_{\mathrm{old}}$, or if
$U_{\mathrm{new}} > U_{\mathrm{old}}$ and $\exp\{(U_{\mathrm{old}} - U_{\mathrm{new}})/kT\}$
is larger than a random number drawn from the range [0,1]. Since the energy is minimised after the proposed
move, the geometric perturbations proposed as steps can generally be much larger than the displacements
used in typical Monte Carlo sampling for thermodynamic properties. In this study we simply used random
perturbations for both the centre of mass positions and orientations of the chiral poles, with amplitudes
$5.0$ (reduced unit) and $2.0$ radians, respectively. We ran $5\times 10^5$ basin-hopping steps for each
set of parameters at $T=2.0$ (in reduced units), unless specified otherwise. All the results
presented here were obtained using our global optimisation program GMIN \cite{GMIN}, which is available
for use under the GNU General Public License.

\subsection{Percolation}

To prevent particles from evaporating during quenches, we enforced a percolating
graph of rod centres, where a path can be constructed
between all particles such that the minimum distance between each pair is less
than a chosen percolation distance.\cite{MorganCDW13} A harmonic
compression was applied to help produce such a structure, which was turned off
once the root-mean-square force was below a chosen cut-off value. Once a percolating structure
had been found, any step that produced a disconnected structure after local
minimisation was rejected. It was therefore necessary to use a sufficiently large
percolation distance for minima with disconnected centres but significant rod/core
interactions to be accepted, but not so large that the chains of rods never interact
significantly during global optimisation. For this system, a percolation distance of 7.5
was found to be sufficient.

\section{Results}

\subsection{Unconstrained System}

Here we present global optimisation results for clusters of chiral rods. We set the length scales
$\sigma_0 = \sigma_r = \sigma_w = 1$, and the energy scales $\epsilon_r = \epsilon_w = 1$, so there are three free parameters:
the chirality angle $\alpha$, the chiral pole strength $\mu$, and the rod length $L$. We focus on
morphologies with $L \gg 1$, where shape anisotropy plays an important role, setting $L=6$ as a
representative example. {\color{black} At small $L$, the rods tend to a spherical shape and the global
minimum structures are similar to standard Lennard-Jones clusters.}\cite{CCD} In fact, the structure of the global minimum is mostly determined by the
rod potential (spherocylinders) for small $\mu$; since this is not the regime of interest here,
we focus on $\mu \geq 2.5$. 

\begin{figure}
\centering
\includegraphics[width = 0.8\textwidth]{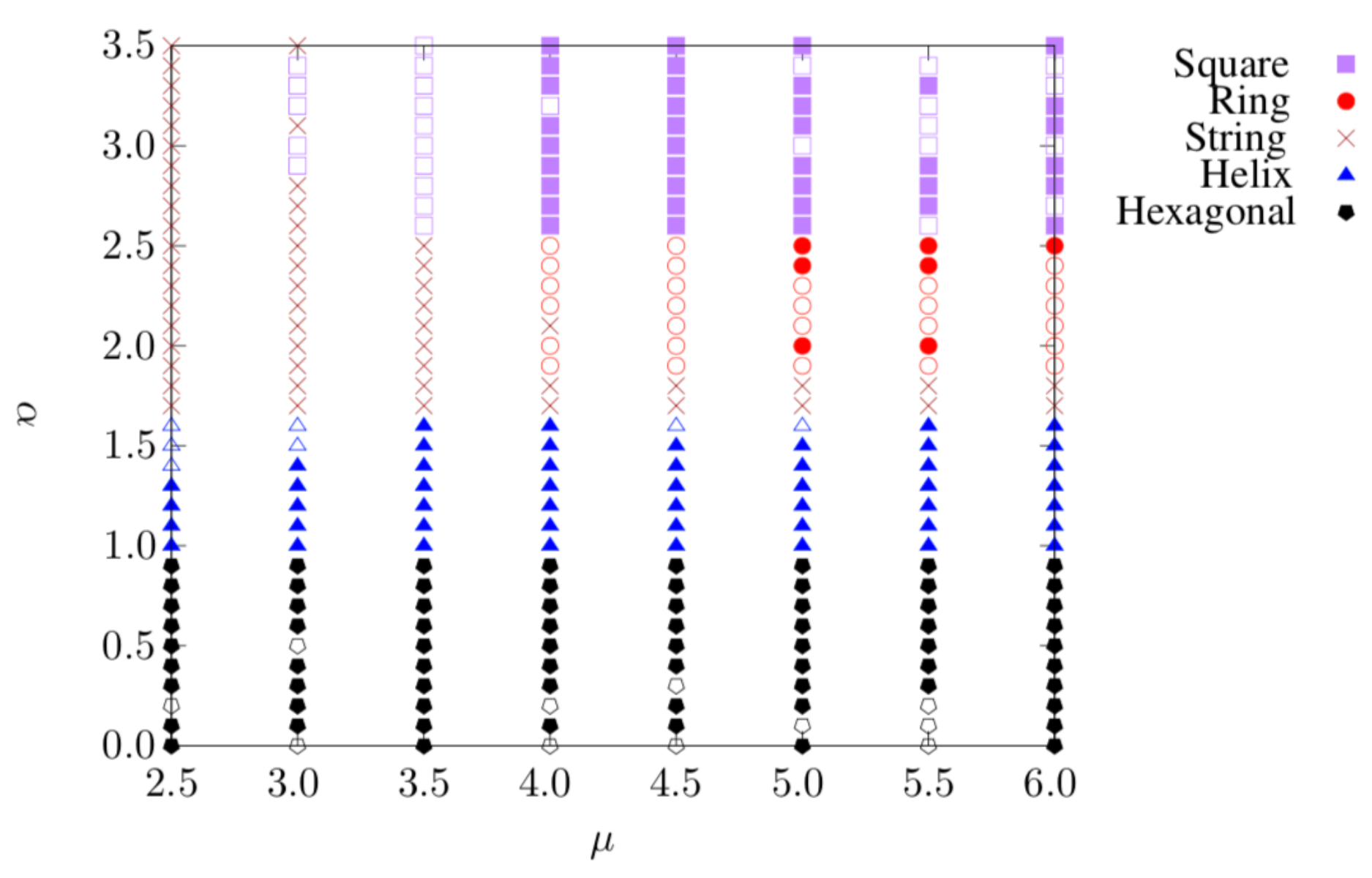}
\caption{The morphology phase diagram as a function of chirality angle $\alpha$ and chiral pole strength
$\mu$, for $N = 25$ rods and $L=6$. Filled symbols indicate points at which three basin-hopping runs converged to
the same structure. Open symbols indicate points where the lowest minimum was not the same in every run. {\color{black}There was
no convergence for any of the string structures.}}
\label{mordiagram}
\end{figure}

The morphology phase diagram for the global minima is shown in figure \ref{mordiagram} as a function of
the chirality angle $\alpha$ and chiral pole strength $\mu$ for clusters of $N = 25$ chiral rods.
We have run global optimisations for other sizes ($N \le 50$) and the results are qualitatively similar.
For each point in parameter space, three basin-hopping runs were started from different random coordinates and run until
all had converged to the same putative global minimum. However, for several values of $\mu$ and $\alpha$,
this condition was not achieved after $3 \times 10^{6}$ basin-hopping steps. In some of these cases, lower energy
minima could be found by starting the search at the coordinates of minima from slightly different
$\mu$ or $\alpha$ that had successfully converged. For parameters where the global minimum was not
consistently located by all three
runs, the lowest energy structure found in any of the runs was taken as the best candidate
for the global minimum.

\begin{figure}
   \includegraphics[width = 0.8\textwidth]{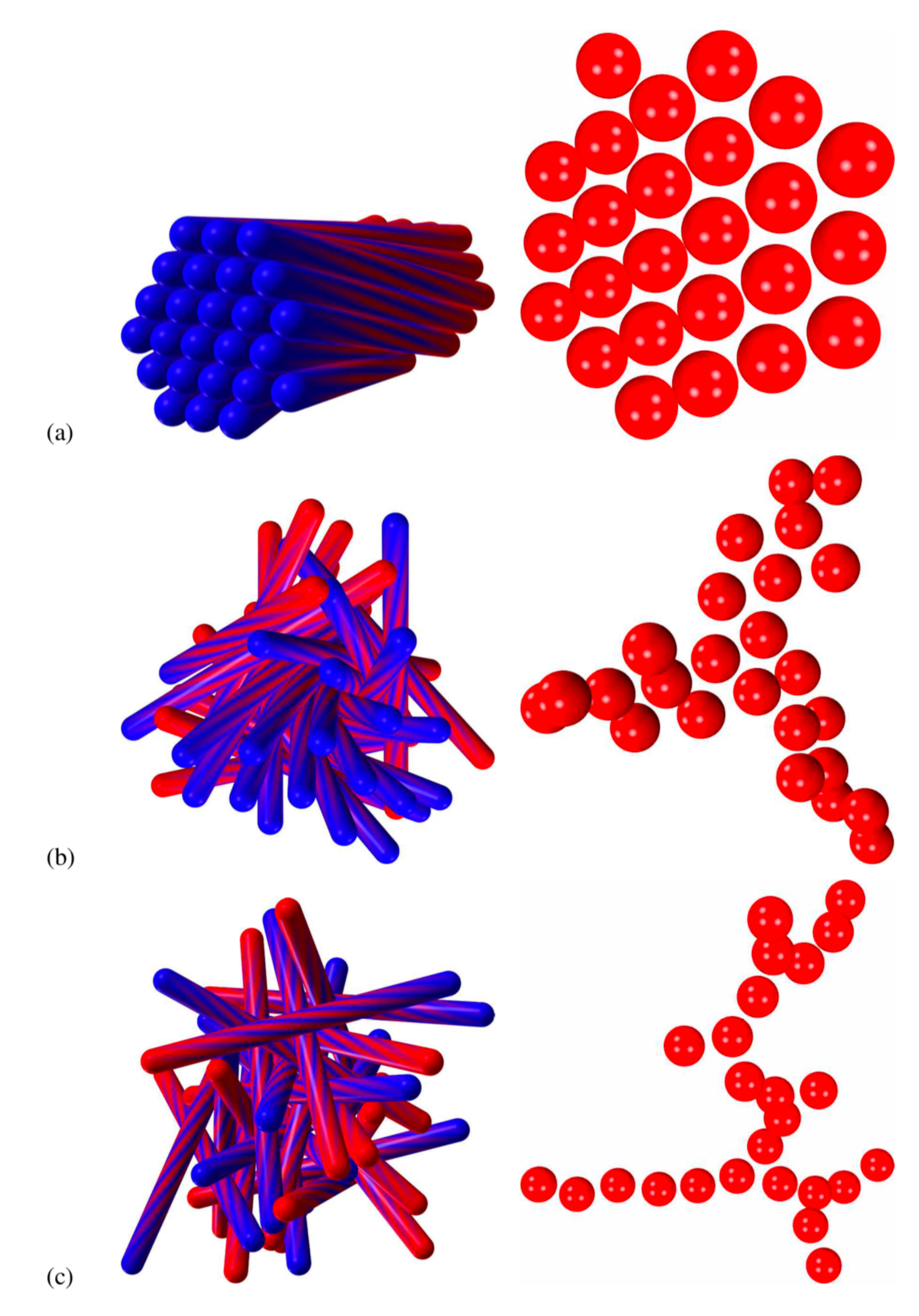}
    \caption{Representative structures for various regions of the parameter space for 25 particles. On the left, the particles are
        represented as rods, with red and blue designating the different ends of the rod. In between, the colours are twisted, with the rate of twist
        proportional to $\alpha$. On the right, the particles are represented as spheres at the centre of mass.
        (a) A hexagonal lattice ($\mu = 4.0$, $\alpha = 0.5$), (b) a helix with three branches ($\mu = 4.0$, $\alpha = 1.3$),
        (c) a branched string-like structure ($\mu = 4.0$, $\alpha = 1.7$). }
    \label{fig:structures1}
\end{figure}

\begin{figure}
   \includegraphics[width = 0.8\textwidth]{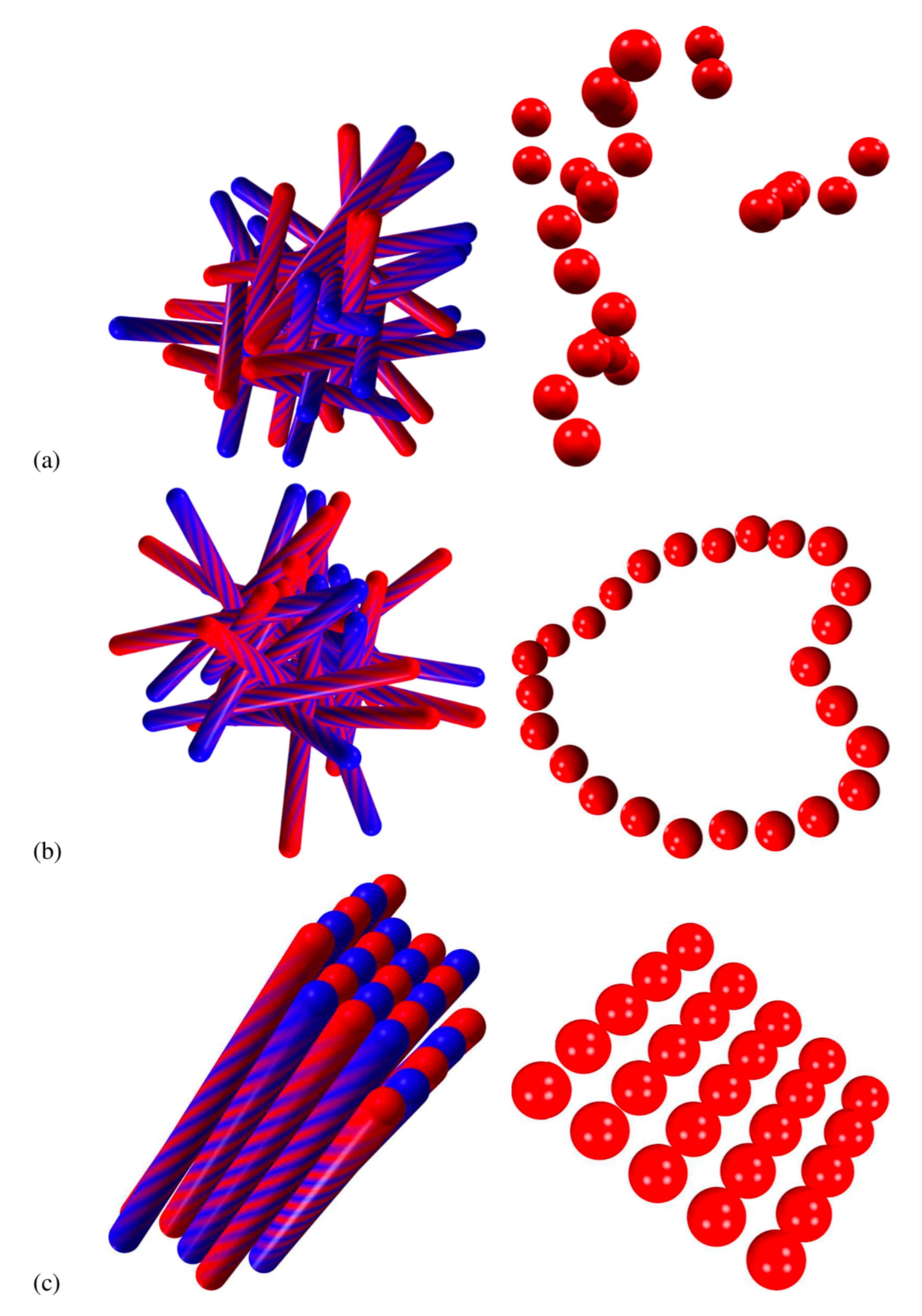}
    \caption{Further representative structures for various regions of the parameter space for 25 particles.  
        (a) A string-like structure broken into multiple chains ($\mu = 2.5$, $\alpha = 2.2$), (b) a ring ($\mu = 4.0$, $\alpha = 2.0$), and
        (c) a square antiparallel lattice ($\mu = 4.0$, $\alpha = 2.7$). }
    \label{fig:structures2}
\end{figure}

The results are {\color{black} generally} insensitive to $\mu$, and instead the main structural
features are determined by variations in the angle $\alpha$. {\color{black} Once $\mu$ is large enough
that the chiral potential becomes the dominant factor, the precise strength has little
effect.} Representative structures
are shown in figures \ref{fig:structures1} and \ref{fig:structures2}; {\color{black} they are all chiral.}

For $\alpha \leq 0.9$ the rod centres
arrange in an approximately hexagonal two-dimensional membrane structure [figure \ref{fig:structures1} (a)], which gives the
closest packing for parallel cylinders. As expected, the chiral interactions between the rods give rise to an overall twist and chirality
(of the same handedness) in the resulting membrane. Reversing the chirality of the interactions
reverses the twist in the hexagonal membrane morphology.
The twisting of the rods away from
parallel increases with $\alpha$ and also towards the edges of the array. In the
central region, the stability gained from closer packing constrains the rods to align
together, whilst at the edges there is more space available for rotations away from
parallel, and so the chiral interactions can be stronger. At $\alpha = 0.9$ the rod centres nearest the
boundary become less ordered and are slightly displaced from the plane of the array,
but the hexagonal symmetry of the central region remains mostly intact.

For $1.0 \leq \alpha \leq 1.6$, the planar structure changes into a helical structure.
When $\alpha = 1.0$ there is still an approximately hexagonal arrangement around each
rod, and then as $\alpha$ increases the number of nearest neighbours decreases. For
$\mu > 4.0$ and $\alpha = 1.2$ or $\alpha = 1.3$, the helix forms three distinct branches, twisted relative to one another [figure \ref{fig:structures1} (b)],
which changes to a single-stranded, twisted ribbon structure, as $\alpha$ increases.
For smaller $\mu$ the number of branches formed is less consistent.

This transition from hexagonal membrane to ribbon/helix morphologies is in fact reminiscent of recent
experimental observations for the {\it{fd}} bacteriophage \cite{GibaudBZHWYBO12}. By varying the temperature of their system, Gibaud et al.\ were able to
modulate the chirality of the virus assembly. At low chirality, a membrane
morphology is stable. However, with increasing chirality,
the perimeter of the membrane undulates, leading to branching of the virus particles
into several arms. Experimental evidence shows that the morphology of each arm corresponds
to a twisted ribbon.
There is a significant difference in the number of
particles between the results presented here and the experimental structures, but
the similarity between the observed morphologies suggests that
the present potential could provide insight into larger chiral structures.

For $1.7 \leq \alpha \leq 2.5$ the rod centres tend to align in a sequential string-like chain [figure \ref{fig:structures1} (c)].
At $\mu > 4$ the ends of this chain tend to meet and form a ring-like structure [figure \ref{fig:structures2} (b)],
which is a challenging target for unbiased global optimisation.
Often low-energy branched structures are easier to find. The ring structure is expected to be more stable, due to the increase in
nearest-neighbour pairs, as well as transannular interactions between
rods pointing across the ring. Generally a sharp decrease in energy appears between
the branched and ring structures. {\color{black} The transition from strings to rings is the only
effect we see due to increasing $\mu$. Larger $\mu$ increases the contribution of the extra nearest-neighbour pairs,
decreasing the energy of the ring structure compared to the string, so making it easier to locate.}

For $\mu \leq 3.5$ no rings were located. In fact,
for $\mu = 2.5$ the most common local minima found were two or three chains of rod
centres, for which the closest intercentre distance is relatively large [figure \ref{fig:structures2} (a)].
Since the chiral interactions of equation \eqref{eq:chiro_pot}
are calculated between the centres of each pair of rods, we expected that a
connected structure of rod centres would be lowest in energy. However, at small $\mu$ the
the rod-rod interactions can be competitive with the chiral interactions, in which case
the global minimum structure may become a more complex three-dimensional
arrangement of interspersed rods. 

For $2.6 \leq \alpha \leq 3.5$ and $4.0 \leq \mu \leq 6.0$ a planar ordered array of rod centres was
again observed, this time in a square lattice with an antiparallel orientation between
nearest neighbours [figure \ref{fig:structures2} (c)]. The square arrangement allows all sets of nearest-neighbour
pairs to be aligned antiparallel, which is not possible in the closer-packed hexagonal
structure. {\color{black} This square arrangement is only possible due to the unsymmetrised nature of the potential, as it
requires $\alpha$ to approach $\pi$.} As for the hexagonal structure, twisting is most significant
at the edges. As $\alpha$ increases towards $\pi$, the twisting decreases, and for $\alpha > \pi$ the
direction of twisting in the array is reversed. For example, for $\mu = 5.0$, the structures
for $\alpha = 2.8$ and $\alpha = 3.5$ are similar both in morphology and energy, but with
opposite chirality.

For $\mu = 3.5$ and $2.6 \leq \alpha \leq 3.5$ a similar antiparallel arrangement was observed,
although the planar array was incomplete, with linear chains positioned below or
above the lattice, or with multiple fragments of antiparallel lattice. The incomplete structure
has fewer second-nearest neighbour pairs, for which there is a repulsive
chiral interaction, but a larger number of favourable rod-core interactions
involving the particles below the plane and the ends of the rods in the array. As $\mu$
decreases, it appears that the nearest-neighbour chiral interactions are insufficient
to maintain a complete $5 \times 5$ square arrangement. The same phenomenon is not
observed for the hexagonal phase, probably due to better packing and favourable chiral
interactions between all pairs of rods.

{\color{black} The system was also studied for $L = 12$ and $L = 24$ with the same
parameters as the representative structures in the figures. Apart from minor
differences in energy, two significant trends were found. Longer rods disfavour
rings, as it is more difficult to arrange the favourable transannular interactions.
Strings were observed rather than rings at $\mu = 4.0$, $\alpha = 2.0$. Also,
longer rods decrease the maximum twist supportable in the square lattice. For
$\mu = 4.0$, $\alpha = 2.7$, at $L = 12$ a broken lattice with some square
regions was observed, while a string structure was found for $L = 24$. }

\subsection{Constrained System}

Further basin-hopping runs were carried out with the system constrained in a cylinder. A
cluster of 61 particles can form a symmetric hexagonal disc, so we carried out
runs with 244 particles, starting from four hexagonally arranged layers.
The angles were initially randomly distributed within 0.15 radians of alignment
with the positive $z$ direction. {\color{black} Jars with radii of $R = 4.75$, $R = 5.25$ and $R = 5.75$ were modelled.
The first of these values is a tight constraint with little freedom for the particles to twist. Higher values
of $R$ allow more twisting.}
These radii are large enough to preserve the approximate hexagonal layered arrangement, but small
enough to influence the packing. The jar height was chosen as $h = 55$; small
enough to preserve the initial layered arrangement, while large enough to allow
some translational freedom. We selected $\mu = 5$ and values of $\alpha$ between
$0$ and $1.7$, where the chiral interaction for the chiral interaction to have a significant
influence on the structure.

\begin{figure}
    \centering
   \includegraphics[width = 0.8\textwidth]{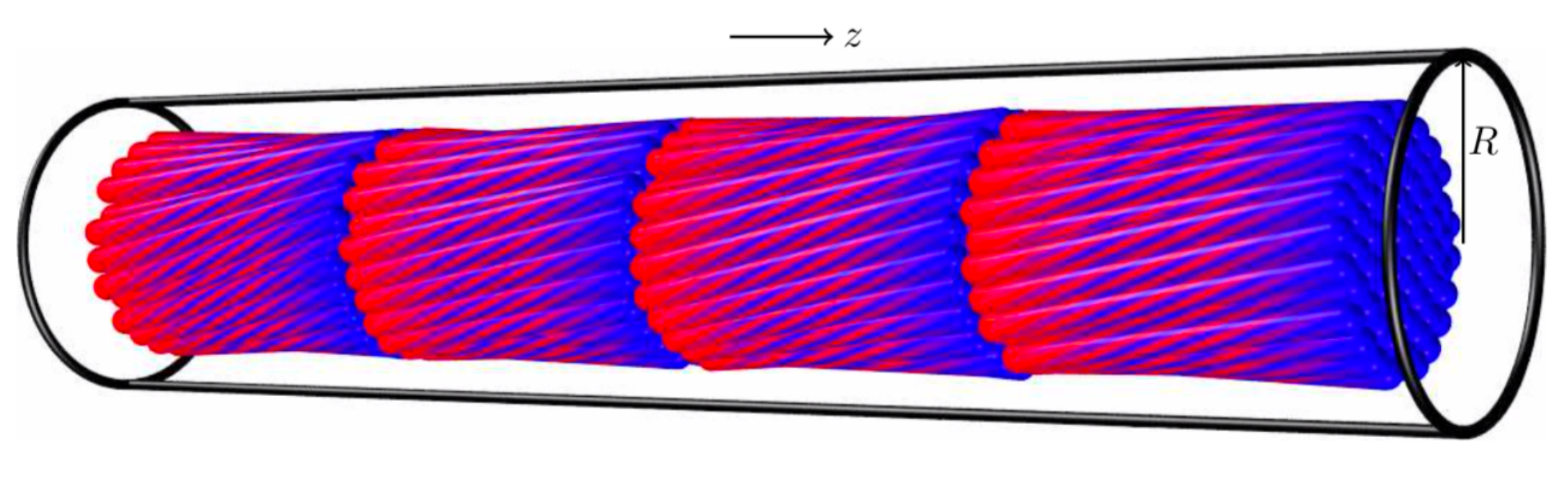}
    \caption{The lowest energy structure located for 244 particles constrained in a jar,
        with $\mu = 5$, $\alpha = 1.5$ and $R = 4.75$.
        The jar is shown in outline; rods are represented as
        in figures \ref{fig:structures1} and \ref{fig:structures2}.}
    \label{fig:jar}
\end{figure}

The lowest energy structure found for the system with $\alpha = 1.5$ {\color{black} and $R = 4.75$} is shown in figure \ref{fig:jar}.
The four layers are maintained, with
approximately hexagonal packing and each layer shows a distinct twist. Although this is
unlikely to be the true global minimum, we believe it should be representative of the
favoured morphology.

\begin{figure}
    \centering
   \includegraphics[width = 0.8\textwidth]{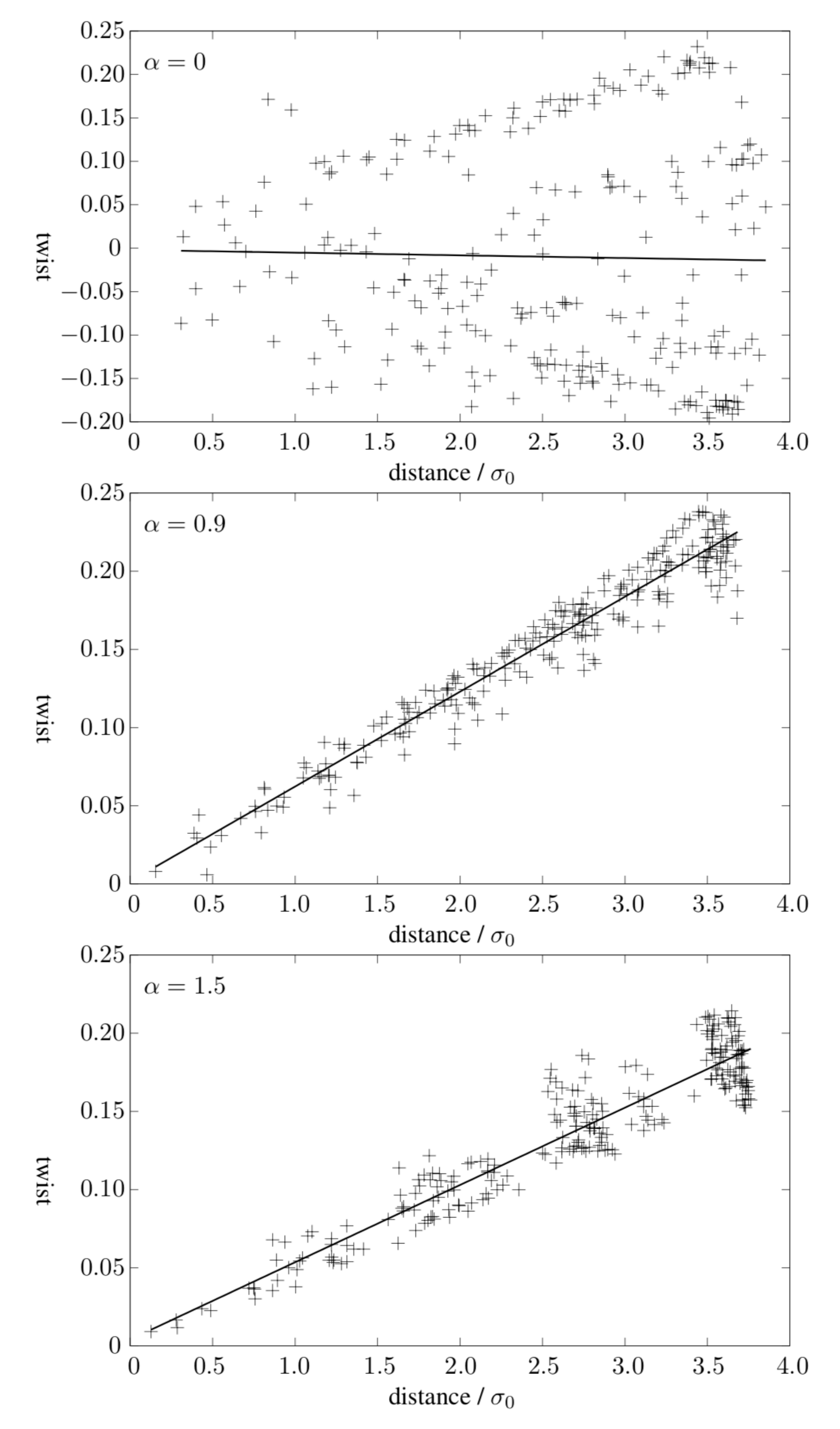}
    \caption{Representative plots for different chiral angles $\alpha$ (in radians) and $R = 4.75$ of the twist of
             each particle against its distance from the container axis. Top:
             $\alpha = 0$; middle: $\alpha = 0.9$; bottom $\alpha = 1.5$.}
    \label{fig:twist}
\end{figure}

To analyse the twist, we have evaluated the projection of a unit vector pointing along
each rod onto a unit vector
perpendicular to the vector connecting the rod centre of mass and the axis of the
cylindrical container. For the lowest energy structure found at each value of
$\alpha$, we have plotted these projections against the distance of the particle
from the container axis, as shown in figure \ref{fig:twist} for selected values of $\alpha$ with $R = 4.75$. For small $\alpha$,
there is little energetic preference for an overall twist {\color{black} in one direction} and the particles sample a
range of values. At moderate $\alpha$, there is some preference for a layer to
organise with an overall twist, but a disordered or oppositely twisted layer is also possible in
individual layers, with a small increase in energy. At larger $\alpha$,
the twisted layers become more energetically favourable, with a more regular hexagonal ordering
within the layer. Comparing $\alpha = 0.9$ and $\alpha = 1.5$ in figure \ref{fig:twist}, the
clustering of points for higher $\alpha$ is due to a greater degree of hexagonal ordering.
Beyond $\alpha = \pi / 2$, the structure becomes disordered, with the layer organisation
breaking down, as aligning particles in similar directions is no longer favourable.

\begin{figure}
    \centering
   \includegraphics[width = 0.9\textwidth]{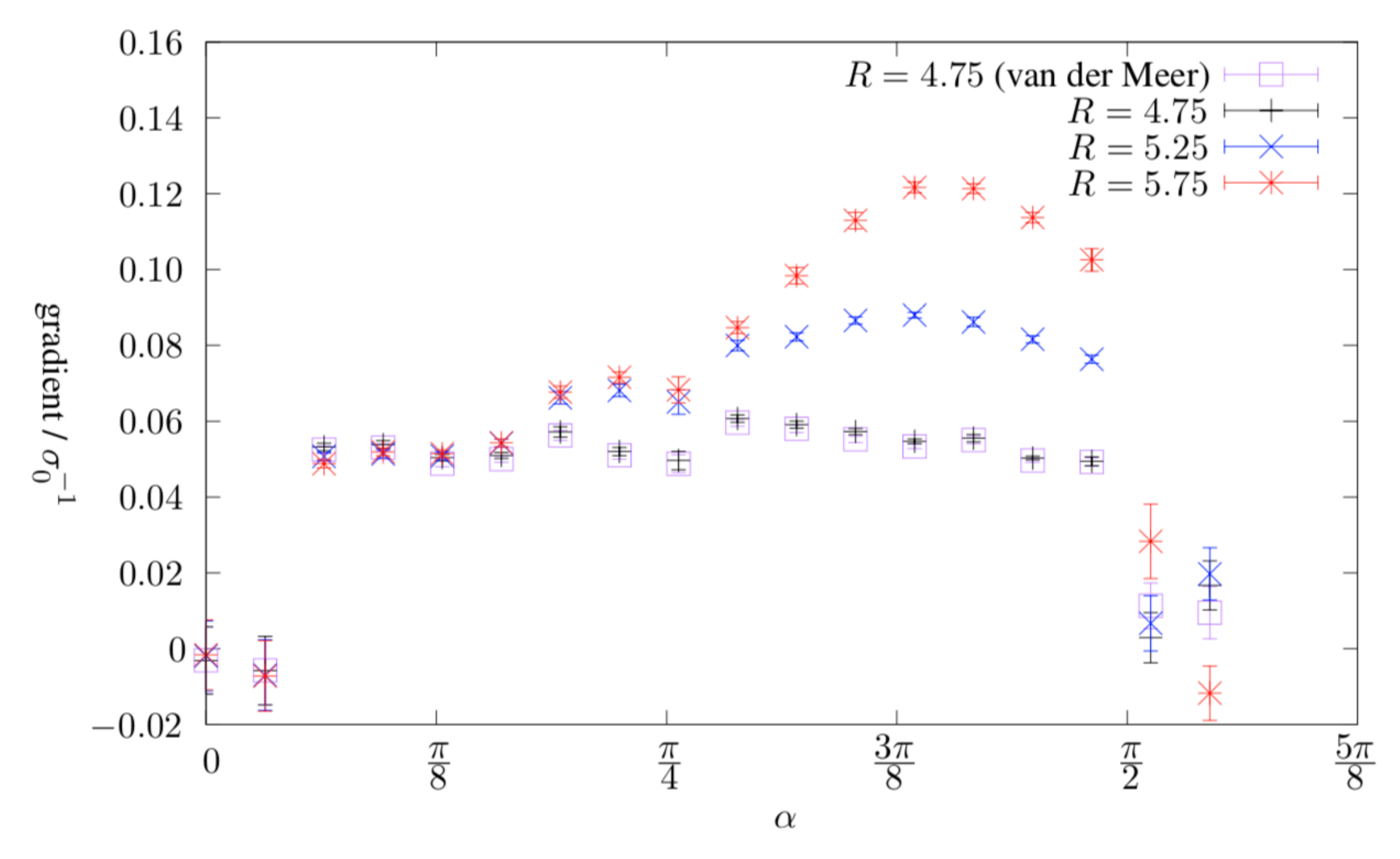}
    \caption{Gradient of the linear best fit for the variation of
        the rod direction against the distance of the particle centre of mass from
        the $z$ axis of the cylindrical container, for the lowest energy structures found {\color{black} at
        different values of $\alpha$. The results for the van der Meer potential with $R = 4.75$ are also shown}. The error bars show the
        asymptotic standard error of the fit.}
    \label{fig:twistfit}
\end{figure}

The gradient of the linear best fit of twist against distance was calculated for each value of $\alpha$, {\color{black} with
$R = 4.75$, $5.25$ and $5.75$}. {\color{black} For higher of $R$, the basin-hopping run was begun by relaxing the lowest energy
structure found for $R = 4.75$ at the same value of $\alpha$. In no cases did the basin-hopping runs locate a structure lower in
energy than the initial relaxed structure.} The values are shown
in figure \ref{fig:twistfit}. The gradient is large for values of $\alpha$ between
$0.2$ and $1.5$, where there is a tendency for the particles to twist, but does not
change significantly with $\alpha$ within this range for $R = 4.75$, {\color{black} since the maximum
twist is tightly constrained by the small radius. At higher $R$, the gradient increases
further as larger twists are compatible with the size of the jar. The layers
have a preferred twist from the chirality of the particles, that may be limited by the jar.}
Beyond $\alpha = \pi / 2$, the particles no longer adopt well-ordered twisted layers.

{\color{black} The simulations with $R = 4.75$ were rerun with the van der Meer potential. Basin-hopping runs were started
from the lowest energy structure found for each value of $\alpha$. The minimum energy
found was that obtained by an initial local minimisation, except for $\alpha = 1.6$. For smaller
values of $\alpha$, the structures and tilts were very similar. At low twisting, there is little
difference between the two potentials. For the disordered structures above $\alpha = \frac{\pi}{2}$, the structures
are qualitatively similar.}

\section{Conclusions}

We have described a single-site potential for simulating chiral interactions, and used it to
investigate the energetically favoured morphologies for chiral rod clusters over a wide range of chiral
parameters, using basin-hopping global optimisation. We found that these structures are predominantly determined
by the angle $\alpha$, which encodes the preferred twist angle for a pair of particles.
We observed (i) hexagonal membrane, (ii) ribbon/helix, (iii) ring, and (iv) square lattice
morphologies for increasing values of $\alpha$.

The transition between hexagonal membrane and ribbon/helix morphologies is particularly interesting,
as it is very similar to results observed for {\it{fd}} bacteriophage,
although the number of particles we have used in this preliminary survey is much smaller than in the experiments.
Since the proposed potential is very convenient in computational terms, one avenue for
future research is to carry out large scale simulations of chiral rods, which will
be presented elsewhere. We also note that previous studies have reported structures similar
to the hexagonal membrane and ribbon/helix morphologies, which suggests that these are generic and robust
structures for chiral assemblies.\cite{YanHE08,YanHE09,GibaudBZHWYBO12}

We have also investigated the behaviour when particles are confined within a cylindrical jar.
Layers of particles adopt an overall twist, dependent on their preferred chiral angle, with
the degree of twisting increasing towards the outside of the jar. These results reproduce the arrangements of
chiral fusilli pasta arranged in a macroscopic jar\cite{SchallerB12} \textcolor{black}{and are very similar for our potential
and the van der Meer potential, demonstrating the value of our potential for use in both microscopic and macroscopic simulations.}

\appendix

\section{Gradients of the Chiral Potential}

We now derive the analytical gradients for the chiral potential, which are required for efficient
geometry optimisations. The derivative with respect to the displacement of the particle centres is
\begin{equation}
\frac{dU^c_{ij}}{d\mathbf{r}_{ij}} = -\frac{\mu^2 \sigma_0^3}{r_{ij}^4}
\{ \cos\alpha \left[-3\left(\hat{\bm{\mu}}_i \cdot \hat{\bm{\mu}}_j\right) \hat{\mathbf{r}}_{ij} \right] +
\sin\alpha \left[ \left(\hat{\bm{\mu}}_i \times \hat{\bm{\mu}}_j\right) - 4 \left(\left(\hat{\bm{\mu}}_i
\times \hat{\bm{\mu}}_j\right) \cdot \hat{\mathbf{r}}_{ij} \right) \hat{\mathbf{r}}_{ij} \right] \} \, .
\end{equation}
To compute the angular gradients, we first write
\begin{equation}
\hat{\bm{\mu}}_i = \mathbf{R}_i \hat{\bm{\mu}}^0_i \, ,
\end{equation}
where $\mathbf{R}_i$ is the rotation matrix for chiral particle $i$, and $\hat{\bm{\mu}}^0_i$ is its orientation
in a fixed reference. We use the angle-axis framework \cite{Wales05,ChakrabartiW09,RuhleKCW13} to describe the rotational
degrees of freedom. Here, the vector $\mathbf{p} = \psi \hat{\mathbf{p}}$ describes a rotation by an angle $\psi$
around the axis defined by the unit vector $\hat{\mathbf{p}}$. The derivatives of the chiral potential with
respect to angle-axis components are 
\begin{eqnarray}
\frac{dU^c_{ij}}{dp^x_i} &=& -\frac{\mu^2 \sigma_0^3}{r_{ij}^3}
\{ \cos\alpha \left[\left(\mathbf{R}^x_i\hat{\bm{\mu}}^0_i\right) \cdot \hat{\bm{\mu}}_j \right] +
\sin\alpha \left[ \left(\mathbf{R}^x_i \hat{\bm{\mu}}^0_i \right) \times \hat{\bm{\mu}}_j \right] \cdot \hat{\mathbf{r}}_{ij} \} \, , \\
\frac{dU^c_{ij}}{dp^x_j} &=& -\frac{\mu^2 \sigma_0^3}{r_{ij}^3}
\{ \cos\alpha \left[\hat{\bm{\mu}}_i \cdot \left(\mathbf{R}^x_j\hat{\bm{\mu}}^0_j\right) \right] +
\sin\alpha \left[ \hat{\bm{\mu}}_i \times \left(\mathbf{R}^x_j \hat{\bm{\mu}}^0_j \right) \right] \cdot \hat{\mathbf{r}}_{ij} \} \, ,
\end{eqnarray}
where $x$ signifies one component of the angle-axis vector and $\mathbf{R}^x_i$ is the derivative of the
rotation matrix for particle $i$ with respect to the $x$ component of the angle-axis vector.

\section{Spherocylinders}

Two rods of half-length $L$ and $L$ centred at $\mathbf{r}_{i}$ and $\mathbf{r}_{j}$ with
orientations defined by their poles $\hat{\bm{\mu}}_{i}$ and $\hat{\bm{\mu}}_{j}$ have
a distance of closest approach
\begin{equation}
    d = \min_{\mathbf{x}_{i} \in S_{i}, \mathbf{x}_{j} \in S_{i}} \lvert \mathbf{x}_{i} - \mathbf{x}_{j} \rvert,
\end{equation}
where $S_{i}$ is the set of all the points in rod $i$. It is convenient to write the points of closest
approach $\mathbf{x}_{i}$ and $\mathbf{x}_{j}$ as
\begin{equation}
    \mathbf{x}_{i} = \mathbf{r}_{i} - \lambda_{i} \hat{\boldsymbol \mu}_{i},
\end{equation}
where $-L \leq \lambda^{i} \leq L$. Thus,
\begin{equation}
    d = \min_{\lvert \lambda_{i} \rvert \leq L,\lvert \lambda_{j} \rvert \leq L}
    \lvert \mathbf{r}_{ij} - \lambda_{i} \hat{\bm{\mu}}_{i} + \lambda_{j} \hat{\bm{\mu}}_{j}\rvert.
\end{equation}
The pairwise energy, in terms of $d$, is given in equation \eqref{eq:spherocylinder}. The values of
$\lambda_{i}$ and $\lambda_{j}$ are necessary for computing the energy and gradients, and there is a
deterministic algorithm for finding them:\cite{VegaL94}
\begin{enumerate}
\item Check if $\hat{\bm{\mu}}_{i}$ is parallel to $\hat{\bm{\mu}}_{j}$. If the rods are parallel
and exactly side-by-side, set $\lambda_{i} = \lambda_{j} = 0$. If the two rods are parallel, but
not side-by-side, set $\lambda_{i} = \pm L$, where the sign is the one that places $\mathbf{x}_{i}$
nearer the interior of the other rod, and set $\lambda_{j}$ to the value that chooses the correct contact
point in $S_{j}$.
\item If the two rods are not parallel, compute
\begin{align}
    \lambda_{i} =& \left[1 - \left(\hat{\bm{\mu}}_{i} \cdot \hat{\bm{\mu}}_{j}\right)^{2}\right]^{-1}
    \left[\mathbf{r}_{ij} \cdot \hat{\bm{\mu}}_{i} - \left(\hat{\bm{\mu}}_{i} \cdot \hat{\bm{\mu}}_{j}\right)
    \left(\mathbf{r}_{ij} \cdot \hat{\bm{\mu}}_{j}\right)\right] \\
    \lambda_{j} =& \left[1 - \left(\hat{\bm{\mu}}_{i} \cdot \hat{\bm{\mu}}_{j}\right)^{2}\right]^{-1}
    \left[-\mathbf{r}_{ij} \cdot \hat{\bm{\mu}}_{j} - \left(\hat{\bm{\mu}}_{i} \cdot \hat{\bm{\mu}}_{j}\right)
    \left(\mathbf{r}_{ij} \cdot \hat{\bm{\mu}}_{i}\right)\right].
\end{align}
\item If $\lambda_{i}$ is outside the permitted range, change it to the closest of the two values
$\pm L$. Recompute $\lambda_{j}$ using this $\lambda_{i}$ as input.
\item If $\lambda_{j}$ is outside the permitted range, change it to the closest of the two values
$\pm L$. Recompute $\lambda_{i}$ using this $\lambda_{j}$ as input. If $\lambda_{i}$ is still
outside the permitted range, change it to the closest endpoint.
\end{enumerate}

We note that there is a discontinuity in the gradient (but not the potential itself) when the rods are parallel or
antiparallel. This effect can be traced to the discontinuity in the position of closest approach as the rods are
perturbed from a parallel or antiparallel configuration. Since this discontinuity leads to instabilities in global
optimisation, we smooth the cusp in the gradient by introducing a correction potential of the form
\begin{equation}
\epsilon \left( \frac{1-(\hat{\bm{\mu}}_i \cdot \hat{\bm{\mu}}_j)^2}{\delta} - 1 \right)^\gamma \, .
\end{equation}
The results are not sensitive to $\epsilon$ and $\gamma$ if suitably large values are chosen.
We use $\epsilon = 10^{20} \epsilon_{r}$ and $\gamma = 100$. We also choose $\delta = 10^{-3}$, so that this
correction plays a role only when $\theta < 2^\circ$.

\section*{Conflicts of Interest}
There are no conflicts to declare.

\begin{acknowledgments}
SWO gratefully acknowledges the financial support of the Dr Herchel Smith Fellowship provided by
Williams College, Williamstown, MA. This work was also supported by the EPSRC and the ERC.
Data may be accessed at http://doi.org/10.5281/zenodo.3407225

Example input and output is available at the GMIN web site in the Cambridge Energy Landscape Database
(http://www-wales.ch.cam.ac.uk/CCD.html). GMIN is available for use under the GNU General Public Licence.
\end{acknowledgments}

\def\acsnano{ACS Nano\xspace}
\def\acp{Adv. Chem. Phys.\xspace}
\def\compchem{Comput. Chem.\xspace}
\def\cshpb{CSH Perspect. Biol.\xspace}
\def\farad{Faraday Discuss.\xspace}
\def\eurjbiochem{Eur. J. Biochem.\xspace}
\def\jcp{J. Chem. Phys.\xspace}
\def\jctc{J. Chem. Theory Comput.\xspace}
\def\jmolliq{J. Mol. Liq.\xspace}
\def\jpca{J. Phys. Chem. A\xspace}
\def\liqcr{Liq. Cryst.\xspace}
\def\mclc{Mol. Cryst. Liq. Cryst.\xspace}
\def\molp{Mol. Phys.\xspace}
\def\nat{Nature\xspace}
\def\natmat{Nat. Mater.\xspace}
\def\pccp{Phys. Chem. Chem. Phys.\xspace}
\def\pr{Phys. Rev.\xspace}
\def\prl{Phys. Rev. Lett.\xspace}
\def\pnas{Proc. Natl. Acad. Sci. USA\xspace}
\def\prsa{Proc. Roy. Soc. A.\xspace}
\def\rmp{Rev. Mod. Phys.\xspace}
\def\rsca{RSC Adv.\xspace}
\def\sci{Science\xspace}
\def\softmat{Soft Matter\xspace}
\bibliography{chiro}

\end{document}